\documentclass[preprint,prd,showpacs,showkeys]{revtex4}     
\usepackage{amsmath,graphicx}

\newcommand{\pni}{\par\noindent}
\begin{document}
\title{Traceless stress-energy and traversable wormholes}
\author{A. G. Agnese} 
\affiliation {Dipartimento di Fisica dell'Universit\`a di 
Genova\\Istituto Nazionale di Fisica Nucleare,Sezione di 
Genova\\Via Dodecaneso 33, 16146 Genova, Italy} 
\author {M. La Camera}
\email{lacamera@ge.infn.it} 
\affiliation {Dipartimento di Fisica dell'Universit\`a di 
Genova\\Istituto Nazionale di Fisica Nucleare,Sezione di 
Genova\\Via Dodecaneso 33, 16146 Genova, Italy}
\begin{abstract}
A one-parameter family of static and spherically
symmetric solutions to Einstein equations with a traceless 
energy-momentum tensor is found. When the nonzero parameter
$\beta$ lies  in the open interval $(0,1)$ one obtains 
traversable Lorentzian wormholes. One also obtains naked 
singularities when either $\beta < 0$ or  $\beta > 1$ and 
the Schwarzschild black hole for $\beta = 1$.
\end{abstract}
\maketitle 
\pni In previous papers [1] we have rewritten the 
exterior and interior Schwarzschild solutions replacing the 
usual radial standard coordinate $r$ with an angular one $\psi$ 
defined by 
\begin{equation}
r = \dfrac{2m}{\cos^2\psi} 
\hspace{2cm} -\, \dfrac{\pi}{2} \leq \psi \leq \dfrac{\pi}{2} 
\end{equation}
when $r > 2m$, and analytically continued to
\begin{equation}
r = \dfrac{2m}{\cosh^2\psi}
\hspace{1.7cm} -\, \infty < \psi <  \infty 
\end{equation}
when $r < 2m$, $m$ representing the gravitational mass of the 
considered body.  This allows to obtain some 
results otherwise less apparent or even hidden in the usual 
coordinate systems. With special emphasis on the interior 
Schwarzschild solution, we introduced the concept of 
quasi-universe, namely an universe deprived of a spherical 
void. Two quasi-universes with the same gravitational mass are 
connected by the Einstein-Rosen bridge (fig.1) which can 
therefore be renamed an extreme wormhole., extreme because it is 
not seen as traversable by a static observer.
\begin{figure}[h]
    \centering
    \fbox{\includegraphics[width=1.6in]{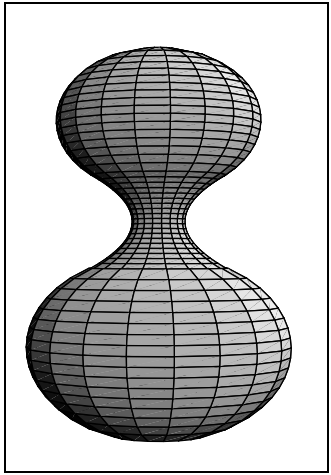}}  
    \caption{The connection between two quasi-universes.}     
\end{figure}
The throat of the Einstein-Rosen bridge contains no matter and by
fact constitutes the $I\cup III$ representation of the exterior
Schwarzschild solution in the customary nomenclature of the 
Kruskal-Szekeres diagram [2]. We interpreted this 
representation as describing two sources of equal gravitational 
mass placed at the boundaries $\psi = \pi/2$ and $\psi = -\, 
\pi/2$, so as a limiting case of fig.1 when the 
connections between the two quasi-universes and the 
Einstein-Rosen bridge tend to infinity.
The quasi-universes can exchange information if the 
Einstein-Rosen bridge is substituted by a two-way traversable 
wormhole. After the seminal paper of Morris and Thorne 
[3] there was a considerable amount of activity about 
wormhole physics [4,5,6,7,8,9,10]. The prescriptions
to have a wormhole traversable are clearly given in [3,4],
and particularly important with this respect is the violation
of the null energy condition (NEC) [11]. In our opinion to
have a wormhole connecting  two quasi-universes it is also
necessary that no singularity lies beyond its throat. This fact 
can occur, as an example, in  the Brans-Dicke theory of 
gravitation when the post-Newtonian parameter $\gamma$ is greater
than unity. In this case one effectively obtains a solution 
describing a wormhole which connects two spaces asymptotically 
flat [7,12], but obtains also another solution where 
beyond the throat there is a singularity smeared on a spherical 
surface asymptotically large but not asymptotically flat 
[12,13,14]. The aim of this paper is to find a wormhole 
solution without singularities beyond the throat and different 
from the Einstein-Rosen bridge only by the occurrence of a 
``vacuum tension'', which implies the presence in the throat of 
exotic matter violating the NEC condition. Having in mind the 
work of Kar and Sahdev [6], where a class of wormhole 
solutions is achieved starting from a choice of the ``potential''
$g_{tt}$ and obtaining $g_{rr}$ as a solution to the constraint 
of vanishing Ricci scalar curvature ($R=0$), we find the metric 
components by assuming a traceless energy-momentum tensor 
($T=0$). Our ansatz implies again $R=0$ but the constraints are 
now on the energy density $\rho$ and on the radial and transverse
pressures $p_{\scriptscriptstyle \parallel}$  and
$p_{\scriptscriptstyle \perp}$: 
\begin{equation}
\left\{ \begin{array}{l}
\rho =0 \\ {} \\
p_{\scriptscriptstyle \parallel}+2p_{\scriptscriptstyle \perp} = 
0 \end{array} \right.
\end{equation}
(Note that sometimes in the literature the radial pressure is 
substituted by minus the radial tension $ \tau$ and the 
transverse pressure is simply denoted by $p$).\pni Let us firstly
consider the static spherically symmetric line element 
\begin{equation}
ds^2 = A(r)dr^2+r^2 d\Omega^2-B(r)dt^2
\end{equation}
where
\begin{equation}
d\Omega^2 = d\vartheta^2+\sin^2\vartheta d\varphi^2
\end{equation}
and the energy-momentum tensor with components
\begin{equation}
T_r^r=p_\parallel(r),\quad T_\vartheta^\vartheta = 
T_\varphi^\varphi = p_\perp(r), \quad T_t^t = -\, \rho(r) 
\end{equation}
The relevant Einstein equations, dropping for simplicity the $r$ 
dependence and denoting by a prime the derivative with respect to
$r$, are in units $c=G=1$
\begin{subequations} 
\label{allequations}  
\begin{flalign}
&\dfrac{B-AB+rB'}{r^2AB} = 8\pi  p_{\scriptscriptstyle \parallel
} \label{equationa} \\  {} \nonumber\\ 
&-\dfrac{1}{4rA^2B^2}\,\bigl\{ 2B^2A'+rA{B'}^2+B\,
\bigl[ \,rA'B'- 2A\,\bigl(\,B' +rB''\bigr) \bigr] \bigr\}
= 8\pi  p_{\scriptscriptstyle \perp } \label{equationb}\\ {} 
\nonumber \\ &\dfrac{A-A^2-rA'}{r^2A^2}= -\,8\pi  \rho 
\label{equationc} \end{flalign}                       
\end{subequations}
Taking into account the constraints (3) one obtains from 
eqs. (7)
\begin{equation} \hspace{-2pt} 
\begin{flalign}
A &= \dfrac{1}{1-\, \dfrac{2\eta}{r}} \\ {} \nonumber \\
B &= \left[ \dfrac{1 + \beta\, \left( 
\sqrt{1-\dfrac{2\eta}{r}}-1 
\right)}{\displaystyle{{}\atop{\alpha}}} \right]^2
\end{flalign} 
\end{equation}
where $\eta$, $\alpha$ and $\beta$ are constants. The constant 
$\alpha$ can be related to the position $r_{obs}$ of the 
observer [1], while $\eta$ and $\beta$ are related to the 
gravitational mass $m$ by comparison with the Newtonian limit at 
large distances. More in detail, if we put 
\begin{equation}
\alpha =    1 + \beta\, \left( 
\sqrt{1-\dfrac{2\eta}{r_{obs}}}-1 \right)  , \quad
\eta = \dfrac{m}{\beta}
\end{equation}
the previous expressions for $A$ and $B$ become
\begin{equation} \hspace{-2pt} 
\begin{flalign}
A &= \dfrac{1}{1-\dfrac{2m}{\beta r}} \\ {} \nonumber \\
B &=  \left[ \dfrac{1 + \beta\, \left( 
\sqrt{1-\dfrac{2m}{\beta r}}-1 \right)}{1 + \beta\, \left( 
\sqrt{1-\dfrac{2m}{\beta r_{obs}}}-1 \right)} \right]^2
\end{flalign}
\end{equation}
so, when the observer is put at spatial infinity, the
asymptotic behaviour of $B$ is effectively $B \approx 1-2m/r$ 
regardless of the sign of $\beta$. As to the pressures, they are
\begin{equation} \hspace{-.1in}
p_{\scriptscriptstyle \parallel} =-\, 2 p_{\scriptscriptstyle 
\perp} = \dfrac{(- 1+\beta )\, m}{4\pi  \beta\, \left[ 1+\beta 
\, \left(\sqrt{1- \dfrac{2m}{\beta r}}-1\,\right)\right]\, r^3}
\end{equation}
and are independent, as they must be, on the observer position 
$r_{obs}$. Therefore once $r_{obs} $ is  fixed in  eq. (12),
the nonzero constant $\beta$ becomes the only free parameter 
in this model. A treatment of the integration constants different
from ours is given in  [15].\pni The invariant of 
curvature $K = 
R_{\alpha\beta\gamma\delta}R^{\alpha\beta\gamma\delta}$, which 
will be needed in the following, is given by 
\begin{equation} 
\begin{flalign} 
K =& \dfrac{24m^2}{r^7 \beta ^2 \left[ 1+\beta 
\, \left(\sqrt{1- \dfrac{2m}{\beta 
r}}-1\,\right)\right]^2}\,\Bigg\{ -\,4m\beta \,+ 
\nonumber \\ {} \\
&r\,\left[1+\beta \,\left( 3\beta - 2\,(-1+\beta)\, 
\sqrt{1- \dfrac{2m}{\beta r}}-2\right)\right]\Bigg\} \nonumber 
\end{flalign} 
\end{equation} 
The lapse function $B$ vanishes at $r_*=2m\beta/(2\beta-1)$ 
when $\beta <0$ or when $\beta \geq 1$. The particular value 
$\beta =1$ describes a black hole in Schwarzschild geometry with 
zero pressures, event horizon at $r_h=2m$, curvature invariant 
given by $48m^2/r^6$ and therefore essential singularity at 
$r=0$. If instead $\beta <0$ or $\beta >1$ both the pressures 
and the curvature invariant $K$ diverge at $r=r_*$ where $B$ 
vanishes so we have no more an event horizon but a naked 
singularity at $r_*$; let us notice that the essential 
singularity occurs now at a value of $r$ greater than the value 
$r_0=2m/\beta$ where otherwise should blow up the function $A$ if
$\beta$ is nonnegative.\pni Let us consider the case $0<\beta<1$.
Now the function $B$ does not vanish and is everywhere finite so
if we introduce the proper length $l$ through 
\begin{equation} 
l=\pm\,\int_{r_0}^r\,\dfrac{dr'}{\sqrt{1-\dfrac{2m}{\beta
r'}}} = 
\pm\, \left\{r\,\sqrt{1-\dfrac{r_0}{r}}+r_0\,\log{\left[ 
\sqrt{\dfrac{r}{r_0}}\,\left(1+\sqrt{1-\dfrac{r_0}{r}}
\right)\right]}\right\}
\end{equation}
and put the spacetime metric (4) in the form
\begin{equation}
ds^2 = dl^2+r^2(l\,)d\Omega^2-e^{2\phi(l)}\,dt^2
\end{equation}
where $\phi(l)=\log{B[r(l)]}$, it is apparent that the above line
element satisfies all the properties required to can describe a 
Lorentzian wormhole with radius of the throat given by $r_0 = 
2m/\beta$ and traversable in principle because of the absence of 
an event horizon. Also the invariant of curvature $K$ and the 
pressures are everywhere  finite, in particular the 
pressures at the throat are given by 
\begin{equation} p_{\scriptscriptstyle 
\parallel} =-\, 2 p_{\scriptscriptstyle \perp} =  
-\,\dfrac{m}{4\pi\beta r_0^3} = -\, \dfrac{\beta^2}{32\pi m^2} 
\end{equation}
Another coordinate patch which covers the entire geometry, 
besides the one given in eq. (16), and which allows to 
obtain somewhat simpler expressions can be obtained if we
replace the standard radial coordinate $r$ not by the proper 
length $l$ but by an angular variable $\psi$, as discussed 
at the very beginning of the paper, which can be given in this 
case by 
\begin{equation}
r=\dfrac{2m}{\beta \cos^2{\psi}}   \hspace{2cm} -\, 
\dfrac{\pi}{2} \leq \psi \leq \dfrac{\pi}{2} 
\end{equation}
The spacetime metric (4) then becomes
\begin{equation}
ds^2 = \left(\dfrac{4m}{\beta\cos^3\psi}\right)^2 d\psi^2 +
\left(\dfrac{2m}{\beta\cos^2\psi}\right)^2 d\Omega^2 -
\left[\dfrac{1+\beta \,(|\sin\psi|-1)}{1+\beta \, 
(|\sin\psi_{obs}| -1)}\right]^2 dt^2 
\end{equation}
and, because of traversability, the observer could even be put 
at the throat where, according to (18), $\psi_{obs}=0$. Therefore
two quasi-universes can be connected by a traversable  wormhole 
though at the expense of violation of the energy conditions. 
\pni Leaving apart discussions about the traversability in 
practice or about known violations of the energy conditions, we 
would like to make some final remarks. We remind that our 
one-parameter solution depends so critically on the parameter 
$\beta$ as to pass at $\beta = 1$ from a naked singularity 
($\beta >1$)  to a wormhole ($0<\beta<1$) through a state 
describing a Schwarzschild black hole ($\beta = 1$). We found an 
analogous  behavior [7] when considering the static 
spherically symmetric vacuum solution of the Brans-Dicke theory 
of gravitation: in that case by varying the post Newtonian 
parameter $\gamma$  different kinds of solutions were obtained 
when passing through the critical value $\gamma =1$. This value 
corresponds to the usual Schwarzschild exterior solution and a 
possible physical interpretation was given [12] by 
introducing spacetime fluctuations at a scale comparable to 
Planck length near the event horizon of a black hole. Another 
explanation, which seems worthy to be further investigated, might
reside on the high nonlinearity of Einstein equations which makes
solutions to change qualitatively, as in the theory of dynamical 
systems, when a reference state loses its stability because a 
``control parameter'', in the actual case $\beta$, passes through
some critical value. 
\begin{acknowledgments} 
This work was completed  after the sudden demise of A.G. Agnese, 
one of the authors, whose long and precious  collaboration is 
gratefully acknowledged. 
\end{acknowledgments} 

\end{document}